\documentclass{physeauth}

\usepackage{graphicx,amsmath,amssymb,color}

\renewcommand\subsubsection[1]{\subsubsection{\underline{#1}}}

\newcommand\be{\begin{equation}}
\newcommand\ee{\end{equation}}
\newcommand\bea{\begin{eqnarray}}
\newcommand\eea{\end{eqnarray}}

\newcommand\ket[1]{\left|#1\right\rangle}
\newcommand\bra[1]{\left\langle#1\right|}

\newcommand\Avg[1]{\langle#1\rangle}

\newcommand\ra{\rightarrow}

\newcommand\re{\Re\textrm{e}}

\newcommand\trm{\textrm}

\newcommand\id{{\rm 1} 
        \hspace{-1.1mm} {\rm I}
        \hspace{0.5mm}}

\newcommand\eq[1]{Eq.~(\ref{#1})}

\newcommand\fig[1]{Fig.~\ref{#1}}

\begin{document}

\begin{frontmatter}

\title{Weak values under uncertain conditions} 

\author[tfp]{Alessandro Romito \thanksref{thank1}},
\author[wi]{Yuval Gefen}

\address[tfp]{Institut f\"ur Theoretische Festk\"orperphysik, 
Universit\"at Karlsruhe, D-76128 Karlsruhe, Germany }
\address[wi]{Department of Condensed Matter Physics, Weizamnn Institute of Science, Rehovot 76100, Israel}

\thanks[thank1]{
Corresponding author. 
E-mail: romito@tfp.uni-karlsruhe.de}

\begin{abstract}
We analyze the average of weak values over statistical ensembles of pre and post-selected states. 
The protocol of weak values, proposed by Aharonov et al.~\cite{Aharonov:1988aa}, 
is the result of a weak measurement conditional on the outcome of a subsequent strong (projective) measurement.
Weak values can be beyond the range of eigenvalues of the measured observable and, in general, can be complex numbers.
We show that averaging over ensembles of pre- and post-selected states reduces the weak value within the range of eigenvalues of $\hat{A}$.
We further show that the averaged result expressed in terms of pre- and post-selected density matrices, allows us to include the effect of decoherence.
\end{abstract}

\begin{keyword}
weak values \sep quantum measurement \sep quantum dots \sep 
charge sensing 
\PACS  \sep 73.21.La \sep 03.65.Ta \sep 76.30. v \sep 85.35.Gv \sep 05.60.-k
\end{keyword}
\end{frontmatter}


\section{Introduction}
\label{introduzione}

Non-invasive quantum measurement has been a long-standing challenge within the framework of quantum mechanics, since such a protocol seems to be incompatible  with the projection postulate~\cite{Neumann:1932aa} in the standard formulation of the quantum measurement.
According to the projection postulate, the measurement in quantum mechanics is a probabilistic process whose outcome is, with a certain probability, one of the eigenvalue of the measured observable.
Immediately following the measurement, the state of the quantum system collapses onto an eigenstate of the measured operator corresponding to the measured eigenvalue.
As opposed to the standard picture, non-invasive (weak) measurements, weakly disturbs the system, while providing only partial information about the state of the latter.
A major step in formulating alternatives to the standard measurement protocol comes from the two state formulation of quantum mechanics, pioneered by Aharonov, Bergamnn and Lebowitz~\cite{Aharonov:1964aa}.
This formalism was originally introduced to discuss the symmetry between past and future during a quantum  measurement, which is explicitly broken by the projection postulate. 
In Ref.~\cite{Aharonov:1964aa} it was shown that the result of a quantum measurement depends symmetrically on both a past state at which the system is prepared (pre-selection) and a future state in which the system is selected following the measurement (post-selection).

Arguably, the most interesting phenomenon in the context of two state formalism is the emergence of weak values (WV) in measurements between pre- and post-selected ensembles~\cite{Aharonov:1988aa,Aharonov:1990aa}.
The protocol for the weak value involves
(i) preselection  of the system in the state $\ket{\chi_0}$;
(ii)  {\it{weak}} measurement of $\hat{A}$;
(iii) projective (strong) measurement of a second observable $\hat{B}$:
the result of the weak measurement is kept provided the measurement of $\hat{B}$ detects the system in a specific, preselected state $\ket{\chi_f}$.
The weak value resulting from this procedure,
\be
_f\Avg{\hat{A}}_0=\langle \chi_f |\hat{A}|\chi_0 \rangle /\langle \chi_f | \chi_0 \rangle \, ,
\label{valore_debole}
\ee
can be orders of magnitude larger than the standard values~\cite{Aharonov:1988aa}, negative~\cite{Aharonov:2002aa} (where conventional strong values would be positively definite), or even complex.
Weak values may allow us to explore some fundamental aspects of quantum measurement, including access to simultaneous measurement of non-commuting variables~\cite{Hongduo:2007aa,Jordan:2005aa}, dephasing and phase recovery~\cite{Neder:2007aa}, correlations between different measurements~\cite{DiLorenzo:2004aa,Sukhorukov:2007aa}, and even new horizons in metrology~\cite{Aharonov:1988aa}.
The access to non-commuting variables allows a direct study of entangled particles, possibly a direct access to quantum statistics of identical particles.

The realization of a weak value protocol requires a high level of control of the dynamics of the quantum system at hand and of the process of measurement.
While some aspects of weak values have been explored in optics based experiments~\cite{Pryde:2005aa}, the desired
high degree of control is now at hand in quantum solid state devices. 
Indeed, the arena of mesoscopic solid state offers very rich physics to be studied through weak values, as well as the possibility of fine tuning and controlling the system's parameters through electrostatic gates and applied magnetic fields. 
In fact, the study of weak values and their implementation in the context of solid state system has been initiated very recently~\cite{Romito:2008aa,Williams:2008aa} and has led to proposal to explore previously unexamined aspects 
of WVs, like their complete tomography~\cite{Shpitalnik:2008aa}

Given that weak values are average values, i.e. are obtained by averaging over many repetitions of an experiment, they will display some noise. 
Such a noise will be two-fold: 
(i) noise of the weak value keeping fixed pre- and post-selected states at a given replica of the experiment, of purely quantum mechanical origin~\cite{Aharonov:2005aa,DiLorenzo:2008aa}; 
or (ii) noise due to the effect of replica-to-replica fluctuations of the pre- and post-selected states.
The latter is relevant in the analysis of weak values in realistic systems where the pre- and post-selected states are obtained 
by tuning certain system's parameters; in particular, this is expected to be the case in all the reported studies of weak values in optics and solid state based systems.
Moreover a WV protocol essentially relies on the quantum coherence of the system at hand. Fluctuations of the system's parameters necessarily affect the measured WVs.

The present work is devoted to the study of weak values averaged over fluctuations in the pre- and post-selected ensembles.
As we further generalize weak values to include pre and post-selected density matrices, this sets the ground for studying decoherence effects within the WV framework.
In the following we present the derivation of WV in a simple model of system-detector interaction (section~\ref{derivazione}), and discuss a protocol to observe the electron's spin WV in a double quantum dot (section~\ref{esempio}) as a specific example. 
We then present our results concerning the average of weak values and analyze them in the specific example of a double quantum dot (section~\ref{media}). 
We further clarify the relation between averages of weak values on one hand, and the evaluation of WVs for non-pure states on the other hand, and employ it to discuss the manifestation of decoherence effects in weak values.

\section{Weak values}
\label{derivazione}


In an ideal von Neuman measurement, the coupling of a system to a detector is described by the Hamiltonian 
\be
\label{hamiltoniana}
H=H_{\textrm{S}}+H_{\textrm{D}}+H_{\textrm{int}} \, , \,\,\,
H_{\textrm{int}}=\lambda g(t) \hat{p}\hat{A} \, ,
\ee
where $H_{\textrm{S(D)}}$ is the Hamiltonian of the system (detector), and $H_{\textrm{int}}$ is the interaction Hamiltonian. 
Here $\hat{p}$ is the momentum canonically conjugate to the position of the detector's pointer, $\hat{q}$, and
$\lambda g(t)$ ($\lambda\ll1$) is a time dependent coupling constant.  
$\hat{A}=\sum_i a_i \bra{a_i}\ket{a}$ is the measured observable.
We assume for simplicity that the free Hamiltonians of the system and the detector vanish and that $g(t)=\delta(t-t_0)$.
Before the measurement the system is in the state $\ket{\chi_0}$, and the detector is in the state $\ket{\phi_0}$, the latter assumed to be a gaussian wave-packet centered at $q=q_0$, $\ket{\phi_0}=C e^{-(q-q_0)^2/4\Delta^2}$.
After the interaction with the detector the entangled state
of the two is 
\be
\label{stato_entangled}
\ket{\psi}=\exp\{-i\lambda \hat{p} \hat{A} \} \ket{\chi_0}\ket{\phi_0} \, .
\ee
A projection onto the state $\ket{\chi_f}$ leaves the detector in the state
\be
\label{spostamento}
\ket{\psi}=\ket{\phi_0} -i \lambda \Avg{\chi_f|\hat{A}|\chi_0}/
\Avg{\chi_f|\chi_0} \hat{p}\ket{\phi_0}
\approx
e^{-i \lambda {}_f\Avg{\hat{A}}_0 \hat{p}}
\ket{\phi_0} \, ,
\ee
that corresponds to a shift in the position of the pointer proportional to $ \re[_{\chi_f} \langle A \rangle_{\chi_0} ]$.
Hence the expectation value of the coordinate of the pointer (initially equal to $q_0$) is given by
\be
\langle \hat{q}\rangle=q_0-\lambda \re[_{\psi}\langle
A\rangle_{\phi}] \, .
\ee
 Under more general conditions (i.e., $\hat{p}$, $\hat{q}$ are not canonically conjugated), the imaginary part of the weak value may be meaningful too~\cite{Steinberg:1995aa}.

We note that the approximation in \eq{spostamento} is valid if $\Delta \ll \max_{i,j} |a_i-a_j|$. This means that
the initial detector's wave function and the shifted one are strongly overlapping.
In fact, in an ideal strong measurement, there is a one-to-one correspondence between the observed value of the detector's coordinate,  $q_{\alpha}$, and the state of the system, $\ket{\alpha}$. 
Within a weak measurement procedure the ranges of values of $q$  that correspond to two distinct states of the system, $\ket{\alpha}$  and $\ket{\alpha'}$,  are described by two probability distribution functions, $P_{\alpha}(q)$  and $P_{\alpha'}(q)$  respectively. 
These distributions strongly overlap. Hence the measurement of $q$ provides only partial information on the state of S.

\section{Weak values of spin in a double quantum dot}
\label{esempio}
 


As a reference model we consider a recent proposal for observing weak values in a solid state device~\cite{Romito:2008aa}.
The device consists of two electron residing on a double quantum dot, a setup that has been recently considered as a candidate for a qubit~\cite{Petta:2005aa}.
A weak value protocol can be realized in this system by means of the (experimentally demonstrated) coherent control
of electrons spin and charge distributions in the dots, and the weak coupling to quantum point contact to be employed as a detector.

\begin{figure}[ht!]
\begin{center}
\includegraphics[width=70mm]{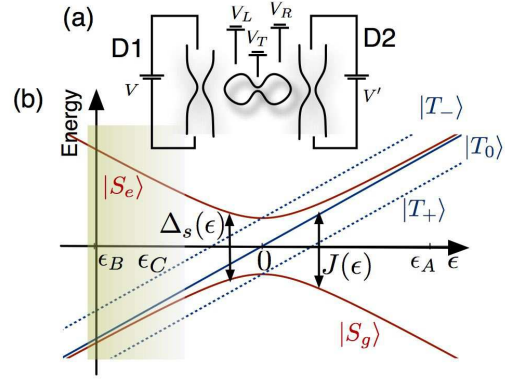}
\end{center}
\caption{Schematics of the system and energy levels. 
(a) Scheme of a double dot with nearby quantum point contacts (QPCs) as charge sensors.
(b) Energy levels of lowest singlet (red) and triplet (blue) states vs. the detuning parameter $\epsilon$. 
In the $(0,2)$ charge configuration the antisymmetric nature of the electrons wave function implies a singlet ground state. 
The  states  $\ket{T_{\pm}}$ (blue dashed lines), with angular momenta $\pm \hbar$  in the direction of the applied magnetic field, are split by the Zeeman energy. 
The range of $\epsilon$ in which the effect of nuclear interaction is relevant is highlighted by the shadowed part.}
\label{fig_1}
\end{figure}

The system (cf. Fig.~\ref{fig_1}(a)) consists of a gate confined semiconducting double quantum dot hosting two electrons. 
The charge configuration in the two dots, $(n_L,n_R)$, is controlled by the gate voltages $V_L$ and $V_R$. 
In particular, by controlling the dimensionless parameter $\epsilon \propto V_R -V_L$, the charge configuration is continuously tuned between $(0,2)$ and $(1,1)$. 
When the two electrons are in the same dot  $(0,2)$, the ground state is a spin singlet, $\ket{S(0,2)}$; the highly energetic excited triplet states are decoupled.  
For  $(1,1)$  the degeneracy of the triplet states is removed by a magnetic field, $\textbf{B}=B \hat{\textbf{z}}$, applied perpendicularly to the sample's plane (cf. \fig{fig_1}).
Due to the spin-preserving inter-dot tunneling, $\Delta_s(0)/2$ (controlled by the gate voltage $V_T$), the ground state of the system is the charge-hybridized singlet, $\ket{S_g(\epsilon)}$, while the first exited state is $\ket{T_0(1,1)}$.
Their energy difference,  $J(\epsilon)=\Delta_s(0)[\epsilon + \sqrt{\epsilon^2+1/4} ]$, is vanishingly small  at $\epsilon \lesssim \epsilon_B$  (cf. Fig. 1(b)). 
In this case the hyperfine interaction between electrons and the nuclear spin~\cite{Erlingsson:2001aa,Khaetskii:2002aa},
facilitates transitions between these two states. 
For our purpose, the effect of the nuclear spins on the electrons is  described by classical magnetic fields, ${{\bf B}_N}_L$,  ${{\bf B}_N}_R$, resulting in the Hamiltonian  $H_N=g \mu_B ({{\bf B}_N}_R-{{\bf B}_N}_L) \cdot \hat{{\bf z}} \ket{T_0(1,1)}\bra{S(1,1)}+ \trm{H.c.}$.
The Hamiltonian for the lowest energy singlet and triplet states can be then written as
\be
\label{hamiltoniana_riferimento}
H=J(\epsilon)\ket{T_0{1,1}}\bra{T_0{1,1}}+H_{N} \, .
\ee

The state of the system is then controlled by tuning the parameter $\epsilon$ between $\epsilon|_B$ and $\epsilon_A$.
By a variation $\epsilon_B \ra \epsilon_A$ fast on the time scale of the nuclear field coupling, $\Avg{T_0(1,1)|H_N|S(1,1)}$, (''fast adiabatic''), $\ket{S(1,1)}$ is mapped into the ground state, $\ket{S(0,2)}$ and $\ket{T_0(1,1)}$ is unchanged. By a "slow adiabatic" variation, the ground state of the Hamiltonian \eq{hamiltoniana_riferimento}, $\ket{\uparrow \downarrow}$, will be mapped into $\ket{S(0,2)}$. 
The described procedure is in fact mapping different spin states at $\epsilon=\epsilon_B$
into different charge states at $\epsilon=\epsilon_A$, and is referred to as spin to charge conversion.
Notably, by measuring the charge of the final state one is effectively measuring the singlet, triplet (for "a fast adiabatic variation") or the $\ket{\uparrow \downarrow}$, $\ket{\downarrow \uparrow}$ (for the "slow adiabatic variation") component of the initial state at  $\epsilon=\epsilon_B$. 

The detectors (D1 and D2) are two quantum point contacts (QPCs) located near the dots. 
They are charge sensors ~\cite{Field:1993aa} suitable for continuous measurements~\cite{Korotkov:2001aa}.
and can be employed both for weak and strong measurement. 
D2 is used to perform an effective strong measurement of the spin via a spin-to-charge conversion
followed by a strong measurement of charge.
By contrast, at $\epsilon \approx \epsilon_B$, the charge difference between the two spin states is 
small and a measuring charge with D1 corresponds to a weak spin measurement.
Formally, the interaction between the double QD and the QPC is modeled as $H_{\trm{int}}=H_{(1,1)} \mathop{P}_{(1,1)}+H_{(0,2)}\mathop{P}_{(0,2)}$, where $\mathop{P}_{(n_L,n_R)}$ is the operator projecting onto the subspace with charge configuration $(n_L,n_R)$.
$H_{(1,1)}$  describes scattering of the electrons in the QPC with transmission (reflection) coefficient $t_0$  ($r_0$): any incoming electron in the QPC, $\ket{\trm{in}}$, evolves to  $\ket{\phi}=t_0\ket{t}+r_0 \ket{r}$, where $\ket{t}$ and $\ket{r}$  are the reflected and transmitted states for the electron.
When the charge configuration in the double dot is $(0,2)$, the QPC is described by the Hamiltonian $H_{(0,2)}$, corresponding to $\ket{\trm{in}}$, evolves to  $\ket{\phi}=t'\ket{t}+r' \ket{r}$.
For $\epsilon \approx \epsilon_B$, the interaction Hamiltonian can be written as $H_{\trm{int}} \approx H_{(1,1)}+ J(\epsilon)/\Delta_s(\epsilon) (H_{(0,2)}-H_{(1,1)})\otimes(\id-\hat{S}^2/2)$, where the measured observable,  $ \hat{A} \equiv \id - \hat{S}^2/2 = \ket{S_g(\epsilon)} \bra{S_g(\epsilon)}$, is the singlet component of the spin state. 
Now, if the QD is in $(1,1)$, the Hamiltonian in the QPC is still $H_{(1,1)}$, while, If the system is in the  $\ket{S_g(\epsilon)}$ state, the electron in the QPC evolves according to 
$\ket{\trm{in}} \ra \ket{\phi'}=(t_0+\delta t(\epsilon))\ket{t}+(r_0+\delta r(\epsilon)) \ket{r} = \ket{\phi}+\ket{\Delta \phi}$. 
$\delta t$, $\delta r$ can be tuned to be arbitrarily small in  $J(\epsilon)/\Delta_s(\epsilon)$.

The protocol for a weak value is realized a sequence of voltage pulses as described in Fig. 2(a). 
\begin{figure}[h!]
\begin{center}
\includegraphics[width=70mm]{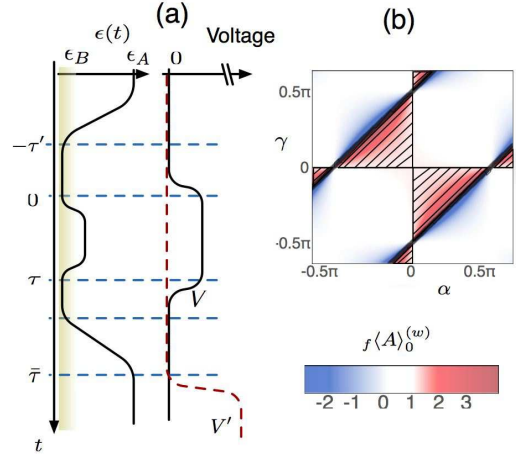}
\end{center}
\caption{ (a) A protocol to measure weak values of two electron spin: shown are  $V$, $V'$,  (voltage bias across D1 and D2 respectively) and $\epsilon$.
(b) The weak value, ${}_f\Avg{\hat{A}}_0^{(W)} =[1-{}_f\Avg{\hat{S}^2}_0^{(W)}]/2$, as a function of the parameters $\alpha$ and $\gamma$  for  $\beta=\pi$. 
The dark region defines the range of parameters for which a positive post-selection is obtained with probability  $P_{(0,2)}<0.5\% $.  
The shadowed region (parallel lines) corresponds to the values of the parameters for which  ${}_f\Avg{\hat{S}^2}_0^{(W)}<0$. 
}
\label{fig_2}
\end{figure}
The evolution of the system in the absence of the detector for this protocol has already been realized in experiment~\cite{Petta:2005aa}. 
{\it pre-selection}:
Initially, at  $\epsilon=\epsilon_A$, the system is in the ground state,  $\ket{S(0,2)}$. 
By a fast adiabatic variation (cf. Fig. 1) it is evolved into $\ket{S(1,1)}$ ($\epsilon=\epsilon_B$ at time $t=-\tau'$). 
This state evolves under the influence of the nuclear interaction until time  $t=0$, 
thus preselecting  $\ket{\chi_0}=\cos \alpha \ket{S_g}-i \sin \alpha \ket{T_0}$ with
 $\alpha=g \mu_B ({{\bf B}_N}_R-{{\bf B}_N}_L) \cdot \hat{{\bf z}} \tau'$. 
The evolution of the system during the measurement pulse, $\mathcal{U}(\tau,0)= \ket{S_g}\bra{S_g}+ \exp(-i \beta) \ket{T_0}\bra{T_0}$, with $\beta=J(\epsilon)\tau$, is reabsorbed to define the effective preselection state at $t=\tau$,
 \be
 \label{preselezione}
 \ket{\chi_0'}=\cos \alpha \ket{S_g}-i e^{i \beta \tau}\sin \alpha \ket{T_0} \, .
 \ee
 {\it weak measurement}: 
The interaction of the system and the QPC creates an entangled state at time $t=\tau$, $\ket{\psi}=\mathcal{U}(\tau,0)\ket{\chi_0}\ket{\phi}+ \hat{A} \mathcal{U}(\tau,0) \ket{\chi_0}\ket{\Delta\phi}$.
 {\it post-selection}:
The following evolution during the time interval $(\tau, \tau+\tau'')$ is governed by the nuclear interaction and that from $\tau+\tau''$ to $\bar{\tau}$  is a fast adiabatic variation. 
At this point the state $\ket{S(0,2)}$ is post-selected by the detector D2.
The evolution  $\mathcal{U}(\bar{\tau},\tau)$ defines the effective post-selected state at time $t=\tau$, $ |\chi_f' \left. \right\rangle = \mathcal{U}^{-1} (\bar{\tau},\tau) \ket{S(0,2)}=\cos \gamma \ket{S_g}+ i \sin \gamma \ket{T_0}$,
 where  $\gamma=g \mu_B ({\bf B}_{NR}-{\bf B}_{NL}) \cdot \hat{{\bf z}} \tau''$ (cf. Fig.~2(b)).
 
Measuring the current in the QPC conditional to the positive outcome of the post-selection of $\ket{S(0,2)}$, gives, at lowest order in $t_0^* \delta t$,
 the weak value of $\hat{A}$,
\bea
\label{valore_dbole_corrente}
& & _f\Avg{I}_0 \approx I_0+(2 e^2V/h)2
 \mbox{\rm Re} \{ {}_f\Avg{\hat{A}}_0^{(W)} t_0^* \delta t \} \, , \\
 & &  _f\Avg{\hat{A}}_0^{(W)}=\frac{\Avg{\chi_f'|\hat{A}|\chi_0'}}{\Avg{\chi_f' |\chi_0'}}=
\frac{ \Avg{\chi_f'|\mathcal{U}(\tau,0)\hat{A}|\chi_0}}{\Avg{\chi_f'|\mathcal{U}(\tau,0) |\chi_0}} \, .
 \eea
Here  $I_0=2 e^2V|t_0|^2/h$ is the current for the $(1,1)$ charge configuration.

With the specific pre- and post-selected states,
the weak value of $\mathcal{U}(\tau,0)\hat{A}$ is then 
\be
\label{risultato}
_f\Avg{\hat{A}}_0^{(W)}=\cos \gamma \cos \alpha /(\cos \gamma \cos \alpha - 
e^{-i \beta} \sin \gamma \sin \alpha) \, .
\ee 
 By tuning the duration of the pulses, one can obtain a real WV (e.g. for  $\beta=\pi$), which is arbitrarily large (e.g. for  $\gamma-\alpha \ra \pi/2$) -- cf. \fig{fig_2}(b). Improvements of the protocol to overcome decoherence effects due to the nuclear spins fluctuations are discussed in Ref.~\cite{Romito:2008aa}.

\section{Weak values averaged over pre- and post-selection}
\label{media}

The model discussed in the previous section 
clearly shows that the weak value depends on
external parameters, namely the angles $\alpha$
and $\gamma$ that control the preselected and 
post-selected state respectively.
The weak value is obtained as an 
average over many replicas of the experiment,
a natural question is then how the weak value is affected by 
averaging over replica-to-replica fluctuations of the 
pre- and post-selected states.
The electron spin weak value case will serve as a reference
to specifically discuss our more general results.

Given that weak value with specific
pre- and post-selected states 
is already obtained as an average over many repetitions
of the measurement, averaging over pre and post-selected states
with distributions $Q_0(\ket{\chi|_0})$,  $Q_{f}(\ket{\chi_f})$ respectively
is obtained through either :
(i) Repeating numerous times the measurement with
 given $\ket{\chi_o}$ and $\ket{\chi_f}$,
keeping only the properly post-selected states, and determining the 
weak value; then iterating the same procedure with different 
pre and post-selected states chosen according their 
distributions, and determining the averaged weak value; or 
(ii) Choosing $\ket{\chi_o}$ and $\ket{\chi_f}$ according to 
the distribution functions $Q_0$, $Q_f$ and performing the 
 weak value protocol {\it once}; then repeating the experiment 
 many times with different states $\ket{\chi_o}$ and $\ket{\chi_f}$ 
 weighted according to the distribution functions $Q_0$, $Q_f$;
then iterating the procedure and averaging the results.

Having in mind the protocol discussed in section \ref{esempio}, 
the former case would correspond to averaging \eq{risultato}
over fluctuations of $\alpha$ and $\gamma$; admittedly it is experimentally
quite artificial. To be specific, consider a case where
 the post-selected state is fixed 
and only the preselected state can fluctuate,
i.e. $\alpha$ fluctuates. The average in (i) corresponds to determining
the average (weighted with $Q_0$) color of \fig{fig_2}(b) along the 
line $\gamma=\trm{cost.}$. 
On the other hand, the definition of averaging according to (ii) 
is the natural result of 
an experiment detecting weak values in the presence of random pre- and 
post-selection.
We discuss the two cases separately. Below we employ the following
 notation for probability distributions: 
$P(a,b)$ is the probability for both the events $a$ and $b$ to occur,
$P(a:b)$ is the conditional probability for $a$ to take place
 given that $b$ did take place,
$P(a,b:c)$ is therefore the probability for the occurrence of
both $a$ and $b$ conditional on the occurrence of $c$.
The following formal relations hold:
\bea
& & P(a,b)=P(a:b) \, P(b)= P(b:a)\, P(a)\\
& & \sum_b P(a,b)=P(b) \\
& & P(a:b,c)=P(a,b,c)/P(b,c)=P(a,b:c)/P(b:c)
\eea

The formal expression for the average defined in (i)
is simply:
\be
{}_{Q_f}\Avg{A}_{Q_0}=\sum_{\ket{\chi_0}} \sum_{\ket{\chi_f}} 
\, Q_0(\ket{\chi_0})Q_f(\ket{\chi_f}) 
{}_{f}\Avg{A}_{0} \, .
\label{mediaA}
\ee
This averaged weak value can lie outside the spectrum of the 
eigenvalues of $\hat{A}$.

In order to properly determine the 
averaged WV defined in (ii),
we start by noting that, quite generally, the weak value can be written
in terms of conditional probabilities as:
\be
\label{probabilita_condizionale}
_{\chi_f}\Avg{\hat{A}}_{\chi_0} =\frac{\sum_i  a_i  
P(a_i, \ket{\chi_f}) }
{ \sum_i  P(\ket{\chi_f},a_i) } =\frac{\sum_i a_i P(a_i,\ket{\chi_f} | \ket{\chi_0})}{\sum_i P(a_i,\ket{\chi_f} | \ket{\chi_0})} \, .
\ee
This equation is easily generalized if the preselected state is chosen from a distribution 
$Q_0(\ket{\chi_0})$,
\be
{}_{\chi_f}\Avg{A}_{Q_0}= 
\frac{\sum_i \sum_{\ket{\chi_0}} a_i Q_0(\ket{\chi_0}) P(a_i,\ket{\chi_f} | \ket{\chi_0})}{\sum_i \sum_{\ket{\chi_0}} Q_0(\ket{\chi_0}) P(a_i,\ket{\chi_f} | \ket{\chi_0})} \, .
\label{due}
\ee
If the measurement of $\hat{A}$ is weak, the initial state is not disturbed by the measurement and we can write
\bea
& & \sum_i P(a_i, \ket{\chi_f} |\ket{\chi_0}) \simeq |\Avg{\chi_f|\chi_0}|^2 \, \label{tre} \\
& & \sum_i a_i P(a_i, \ket{\chi_f} | \ket{\chi_0}) \simeq {}_{\chi_f}\Avg{\hat{A}}_{\chi_0} |\Avg{\chi_f|\chi_0}|^2 \,
\label{quattro}
\eea
where \eq{quattro}  is derived by plugging \eq{tre} into \eq{probabilita_condizionale}. 
Substituting \eq{tre}, \eq{quattro} into \eq{due} we find
\be
{}_{\chi_f}\Avg{A}_{Q_0}=\frac{\Avg{\chi_f | \hat{A} \rho_0 | \chi_f}}{\Avg{\chi_f | \rho_0|\chi_f }} \, .
\label{cinque}
\ee
where $\rho_0=\ket{\chi_0}Q(\ket{\chi_0})\bra{\chi_0}$ 
is the density matrix for the initial state 
given by the distribution $Q_0(\ket{\chi_0})$.
The result in \eq{cinque} is then the generalization of 
weak values to non-pure states. 

If the final state is chosen according
to a distribution function  $Q(\ket{\chi_f})$, 
then the average over such a distribution function 
has to be taken separately for the numerator and denominator.
The previous equation is then modified as
\bea
& & {}_{Q_f}\Avg{\hat{A}}_{Q_0} =\frac{\sum_i  a_i \sum_{\ket{\chi_f}} \, 
 P(a_i, \ket{\chi_f}) Q_f({\ket{\chi_f}})}
{ \sum_i \sum_{\ket{\chi_f}}  P(\ket{\chi_f},a_i) Q_f(\ket{\chi_f})} \nonumber \\
& & =\frac{\sum_i  \sum_{\ket{\chi_f}} \, \sum_{\ket{\chi_0}} \,
a_i Q_0({\ket{\chi_0}}) P(a_i, \ket{\chi_f}:\ket{\chi_0})  Q_f({\ket{\chi_f}})}
{\sum_{i} \sum_{\ket{\chi_f}} \sum_{\ket{\chi_0}} Q_0 (\ket{\chi_0})
P(\ket{\chi_f},a_i:\ket{\chi_0}) Q_ f(\ket{\chi_f})} \, .
\label{generale}
\eea
In the last inequality we include a distribution function, $Q(\ket{\chi_0})$ 
for the initial pre-selected state. 

This expression is quite general, it is based on probability 
considerations and can therefore describe both strong (projective) 
and weak measurements. 
The difference between the two is in the specific form 
of the probability distribution functions.
In the case of a projective measurement of $\hat{A}$, 
$P(a_i, \ket{\chi_f}:\ket{\chi_0})=|\Avg{\chi_f|a_i}|^2|\Avg{a_i|\chi_0}|^2$.
In this case, \eq{generale} reduces to
\be
\label{forte}
_{Q_f}\Avg{\hat{A}}_{Q_0} = _{\rho_f}\Avg{\hat{A}}_{\rho_0}=
\frac{\sum_i a_i \mathop{Tr} [\Pi_{a_i} \rho_0] \mathop{Tr} [\Pi_{a_i} \rho_0]}{\sum_i  \mathop{Tr} [\Pi_{a_i} \rho_0] \mathop{Tr} [\Pi_{a_i} \rho_0]} 
\, ,
\ee
where $\rho_{0}=\ket{\chi_{0}}Q(\ket{\chi_{0}})\bra{\chi_{0}}$ 
and $\rho_{f}=\ket{\chi_{f}}Q(\ket{\chi_{f}})\bra{\chi_{f}}$ 
define the initial and final density matrix respectively, and 
$\Pi_{a_i} \equiv \ket{a_i}\bra{a_i}$.
It appears that  ${}_f\Avg{\hat{A}}_0$ is
within the range of eigenvalues of $\hat{A}$, and it is invariant 
under the exchange of $\ket{\chi_0}$ and $\ket{\chi_f}$.
In the limiting case of uniform distributions, 
the final results are: $ {}_{Q_f=\trm{const}}\Avg{\hat{A}}_{Q_0}=\mathop{Tr}\{\hat{A}\rho_0\}$,
$ {}_{Q_f}\Avg{\hat{A}}_{Q_0=\trm{const}}=\mathop{Tr}\{\hat{A}\rho_f\}$,
$ {}_{Q_f=\trm{const}}\Avg{\hat{A}}_{Q_0=\trm{const}}=\mathop{Tr}\{\hat{A}\}$.

\begin{figure}
\begin{center}
\includegraphics[width=80mm]{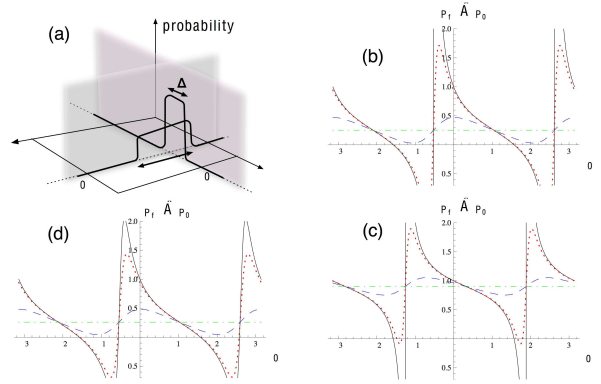}
\end{center}
\caption{Averaged weak values of the electrons spin. (a) Sketch of the distribution functions for the parameters controlling the pre- and post-selected states, $\alpha$ and $\gamma$ respectively. We assume the probability distribution for $\alpha$ to be $P_0(\alpha)=1/(2\Delta)\theta(\alpha+\Delta)\theta(\Delta-\alpha)$ and the probability distribution for $\gamma$  to be $P_f(\gamma)=1/(2\Gamma)\theta(\gamma+\Gamma)\theta(\Gamma-\gamma)$.
(b) Weak value for a preselected ensemble. The plot is the weak value as a function of $\alpha_0$ for $\Delta=0$ (full line), $\Delta=0.25$ (dotted line), $\Delta=1$ (dashed line), $\Delta=\pi$ (dash-dotted line). The post-selection is on a pure state, i.e. $\Gamma=0$ and $\gamma_0=\pi/3$.
(c) Weak values for a post-selected ensemble. The plot is the weak value as a function of $\gamma_0$ for $\Gamma=0$ (full line), $\Gamma=0.25$ (dotted line), $\Gamma=1$ (dashed line), $\Gamma=\pi$ (dash-dotted line). The preselection is on a pure state, i.e. $\Delta=0$ and $\alpha_0=\pi/10$ 
(d) Weak values for both pre and post-slected ensembles. The plot is the weak value as a function of $\alpha_0$ for $\Delta=0$ (full line), $\Delta=0.25$ (dotted line), $\Delta=1$ (dashed line), $\Delta=\pi$ (dash-dotted line). The post-selection is on an ensemble described by the probability distribution $P_f(\gamma)$ with $\gamma_0=\pi/3$ and $\Gamma=0.25$.
}
\label{protocollo}
\end{figure}

In case of a weak measurement, the initial state is unchanged (to
lowest order in the system-detector coupling), 
and we can then assume that
$\sum_{i} P(\ket{\chi_f},a_i:\ket{\chi_0})= P(a_i;\ket{\chi_f}:\ket{\chi_0}) =
|\Avg{\chi_f|\chi_0}|^2$. Also
$\sum_{i} a_i  P(\ket{\chi_f},a_i:\ket{\chi_0})= 
\sum_i a_i P(a_i:\ket{\chi_0},\ket{\chi_f}) P(\ket{\chi_f}:\ket{\chi_0})
={}_f\Avg{\hat{A}}_0 \, |\Avg{\chi_f|\chi_0}|^2 
= \Avg{\chi_f|\hat{A}|\chi_0}\ket{\chi_f|\chi_0}$.
Using these expressions in \eq{generale}, we find
\be
\label{debole}
_{Q_f}\Avg{\hat{A}}_{Q_0}=\mathop{Tr}\{ \rho_f\hat{A}\rho_0\}/\mathop{Tr\{ \rho_f \rho_0\}} \, ,
\ee
with  the same definition of 
$\rho_{0}$, $\rho_{f}$ as in \eq{forte}.
Notably, the weak value undergoes a complex conjugation
operation under the   exchange 
of $\ket{\chi}_0$ and $\ket{\chi_f}$, i.e. time reversal symmetry.
The general behavior is reported in \fig{protocollo};
for uniform distributions we obtain:
$ {}_{Q_f=\trm{const}}\Avg{\hat{A}}_{Q_0}=\mathop{Tr}\{\hat{A}\rho_0\}$,
$ {}_{Q_f}\Avg{\hat{A}}_{Q_0=\trm{const}}=\mathop{Tr}\{\hat{A}\rho_f\}$,
$ {}_{Q_f=\trm{const}}\Avg{\hat{A}}_{Q_0=\trm{const}}=\mathop{Tr}\{\hat{A}\}$.
This demonstrates that if we have incoherent distributions 
for the initial or final state, weak values reduce to 
conventional (strong) values.

As we have already mentioned, the result in \eq{cinque}
is the weak value for a given initial density matrix, obtained 
through probabilistic arguments. We may reproduce the same \eq{cinque}
with the measurement procedure described in section \ref{derivazione},
but replacing the initial pure state by a density matrix, $\rho_0$.
In this case the total density matrix, $R$ of the 
entangled system-detector state after the measurement
is given by 
\begin{equation}
	R=\rho_0 \otimes \ket{\phi_0}\bra{\phi_0} -i \lambda p[\hat{A} \rho_0
 	\otimes \hat{p}\ket{\phi_0}\bra{\phi_0} - \rho_0 \hat{A} \otimes 
	\ket{\phi_0}\bra{\phi_0} \hat{p}] \, .
\end{equation}
Subsequently performing the post-selection, and eventually evaluating the (average) shift of the detector wave packet (to first order in $\lambda$) yields the real part of the weak value.
Within this approach we can discuss the role of {\it decoherence}. 
If we have initially a pure state, but the system is affected by fluctuations leading to decoherence during the weak measurement, it will evolve to a density matrix. If the fluctuations commute with $\hat{A}$ (i.e. fluctuations is described by a term $\xi(t)\hat{B}$ in the Hamiltonian with $[\hat{A}, \hat{B}] =0$) \eq{cinque} still holds replacing $\rho_0$ with $\rho(\tau)$, $\tau$ is the post-selection time.
This is
\be
\label{fine}
_{\chi_f}\Avg{A}_{\rho_0}=\Avg{\chi_f | \hat{A} \rho(\tau) | \chi_f} / \Avg{\chi_f | \rho(\tau)|\chi_f } \, ,
\ee
with $\rho(\tau)= \Avg{\mathcal{U}(\tau,0)\rho_0 \mathcal{U}^{\dag}(\tau,0)}_{\trm{stoc}}$, where $\mathcal{U}(\tau,0)$ is the time evolution operator and the average is 
intended over the the fluctuation of the stochastic parameter.
 
If the fluctuations do not commute with the operator to be measured, a general expression does not exist, 
but perturbation in the coupling to the fluctuating field can be carried out.

\section{Conclusions}

The results presented here clarify the meaning of 
averaging weak values over fluctuations of pre and post-selected states,
which is both a conceptual and experimentally relevant issue.
We have demonstrated that averaging, which results from 
an experimental procedure, is {\it not} simply the average of the weak value 
over the fluctuations of the parameters (as in \eq{mediaA}),  but rather a properly
defined weak value for pre and post-selected density matrices (cf. \eq{debole}).
We have also shown that the derivation of weak values for a density matrix 
sets the basis for incorporating 
decoherence effects into the weak value protocol.

\section{Acknowledgments}

We acknowledge useful discussions with Ya. Blanter and Y. Aharonov.
This work was supported by the Minerva Foundation, the ISF, 
the German-Israeli Foundation (GIF), and the U.S.-Israel Binational
Science Foundation (BSF), and the DFG Priority Programme "Semiconductor Spintronics". AR acknowledge the support of the Alexander von Humboldt Foundation.


\end{document}